\documentclass[a4paper,twoside,reqno]{bjp}
\usepackage{graphicx}
\usepackage{cite}
\usepackage{amssymb,amsmath,amscd,amsthm}
\usepackage{times}

\usepackage[bookmarks=false]{hyperref}
\hypersetup{%
    colorlinks=true,        % false: boxed links; true: colored links
    linkcolor=blue,          % color of internal links (change box color with linkbordercolor)
    citecolor=blue,         % color of links to bibliography
    urlcolor=blue           % color of external links
    }

\usepackage{underscore}

\usepackage{geometry}
\allowdisplaybreaks

\begin{document}

\title{Running Dark Energy and Dark Matter from Dynamical Spacetime}
% Insert the title

% In case the authors are more then three, put the name of the first author followed by the Latin `et al.'

\begin{start}{%
\author{S. Banerjee}{1,2},
% The second argument connects author(s) with addresses
\author{D. Benisty}{2,3,4}
% The second argument connects author(s) with addresses
\author{E. I. Guendelman}{2,3,5}

\address{Inst. for Quantum Gravity, FAU Erlangen-Nuremberg,
Staudtstr. 7, 91058 Erlangen, Germany}{1}
\address{Physics Department, Ben-Gurion University of the Negev, Beer-Sheva 84105, Israel}{2}
\address{Frankfurt Institute for Advanced Studies (FIAS), Ruth-Moufang-Strasse~1, 60438 Frankfurt am Main, Germany}{3}
\address{DAMTP, Centre for Mathematical Sciences, University of Cambridge, Wilberforce Road, Cambridge CB3 0WA, United Kingdom}{4}
\address{Bahamas Advanced Study Institute and Conferences, 4A Ocean Heights, Hill View Circle, Stella Maris, Long Island, The Bahamas}{5}

\received{Day Month Year (Insert date of submission)}
% Insert date of submission
}

\begin{Abstract}
Running Dark Energy and Dark Matter models are candidates to resolve the Hubble constant tension. However the model does not consider a Lagrangian formulation directly. In this paper we formulate an action principle where the Running Vacuum Model (RVM) 
is obtained from an action principle, with a scalar field model for the whole dark components. The Dynamical Spacetime vector field $\chi_\mu$ is a Lagrange multiplier that forces the kinetic term of the scalar field to behave as the modified dark matter. When we replace the vector field by a derivative of a scalar the model predicts diffusion interactions between the dark components with a different correspondence to the RVM. We test the models with the Cosmic Chronometers, Type Ia Supernova, Quasars, Gamma ray Bursts and the Baryon Acoustic Oscillations data sets. We find that $\Lambda$CDM is still the best model. However this formulation suggests an action principle for %$\Lambda$CDM,
the RVM model and other extensions.
\end{Abstract}

\begin{KEY}
Dark Energy; Dark Matter; Modified Gravity.
\end{KEY}
\end{start}

%%%%%%%%%%%%%%%%%%%%%%%%%%%
\section{Introduction}
Almost twenty years after the observational evidence of cosmic acceleration, the cause of this phenomenon, labeled as dark energy remains an open question which challenges the foundations of theoretical physics: The cosmological constant problem - why there is a large
disagreement between the vacuum expectation value of the
energy momentum tensor which comes from quantum field
theory and the observable value of dark energy density \cite{Weinberg:1988cp,Lombriser:2019jia,Frieman:2008sn}. The simplest model of dark energy and dark matter is the $\Lambda$CDM that contains non-relativistic matter and a cosmological constant.

Modification for gravity or to the dark sector were considered in many cases, such as \cite{Anagnostopoulos:2017iao,Saridakis:2018unr,Anagnostopoulos:2018jdq,Basilakos:2019hlb,Anagnostopoulos:2019miu,Vagnozzi:2019ezj,Vasak:2019nmy,Yang:2020uga,DiValentino:2020vnx,DiValentino:2020leo,Benaoum:2020qsi,Yang:2020myd,DiValentino:2020kha,DiValentino:2020evt,Yang:2020tax}. Unification between dark energy and dark matter from an action principle were obtained from scalar fields \cite{Scherrer:2004au,Arbey:2006it,Chen:2008ft,Leon:2013qh,Guendelman:2015rea,Staicova:2018yrc} including Galileon cosmology \cite{Leon:2012mt} or Teleparallel modified theories of gravity \cite{Kofinas:2014aka,Skugoreva:2014ena,Guendelman:2015jii,Guendelman:2016kwj}. A diffusive interaction between dark energy and dark matter was introduced in \cite{Koutsoumbas:2017fxp,Haba:2016swv,Benisty:2017eqh,Benisty:2017rbw,Staicova:2016pfd,Perez:2020cwa}. Interacting scenarios prove to be efficient in alleviating the known tension of modern cosmology, namely the $H_0$ ~\cite{Yang:2018euj,DEramo:2018vss,Yang:2018uae,Guo:2018ans,Kumar:2019wfs,Agrawal:2019lmo,Yang:2019qza,Yang:2019uzo,Benisty:2019vej,Benisty:2019jqz,Benisty:2018gzx,DiValentino:2019jae,DiValentino:2019ffd}. Despite the extended investigation of interacting scenarios the choice of the interaction function remains unknown. 
The Running Vacuum Model \cite{Sola:2016jky,Rezaei:2019xwo,Gomez-Valent:2014rxa,Sola:2015wwa,Sola:2017jbl,Sola:2007sv,Tsiapi:2018she,Basilakos:2015yoa,Farrugia:2018mex,Mavromatos:2015jzo,Basilakos:2019acj,Basilakos:2019mpe,Basilakos:2019wxu,Gomez-Valent:2017idt,Novikov:2016hrc,Novikov:2016fzd} is a good modified model for the cosmological background particularly \textbf{because they can resolve some of the tensions existing in $\Lambda$CDM, mainly the $\sigma_8$ tension. In particular, it was shown in \cite{Rezaei:2019xwo}, that models that include dynamical components of $\dot{H}, H^2$ are far more favoured than $\Lambda$ CDM. Such models predicts a value of $\sigma_8$ between $0.74-0.77$, and that of $H_0$ between local and Planck measurements, thus significantly relaxing both the tensions.} The main point of the RVM comes from Quantum Field Theory (QFT) in a curved spacetime, but here we formulate an action principle that approach the RVM at late times. In this paper we work with Running Dark Energy and Dark Matter from Dynamical Spacetime. Because of the conformal invariance of the radiation it cannot deviate from $1/a^4$, which cannot be modified by going from one conformal frame to another. In fact, running dark radiation is impossible in our Lagrangian framework (or any other Lagrangian framework known so far). Hence, it is probably best to discard this aspect of RVM models.
Other Lagrangian frameworks dealing with RVM also leave out
the radiation (see Appendix in \cite{Sola:2020lba}).

For a homogeneous expanding universe, the RVM expects that the vacuum energy density and the gravitational coupling are functions of the cosmic time through the Hubble rate, assuming the canonical equation of state $p_\Lambda=-\rho_\Lambda(H)$ for the vacuum energy density. The corresponding Friedmann equation (with the presence of radiation $\rho_r$ and pressureless matter density $\rho_m$) reads:
\begin{equation}
H^2/H_0^2 = \tilde{G}(H)\,(\Omega_m+\Omega_r+\Omega_\Lambda(H)),
\end{equation}
where we set $8\pi G = 1$. 
The parameters  $\Omega_{i}^0=\rho_{i0}/\rho_{c0}$ are the current cosmological parameters for matter and radiation. For $i=\Lambda$, it represents the density parameter for vacuum energy. The normalized running Newtonian Constant $G(H)$ is defined as:
\begin{equation}
\tilde{G}(H) := G(H)/G.
\end{equation}
The RVM structure for the dynamical vacuum energy assumes the expansion:
\begin{equation}
\Omega_\Lambda(H;\nu,\alpha)=\Omega_\Lambda^{0}+\nu \left(\frac{H}{H_0}\right)^2+\frac{2}{3}\alpha \frac{\dot{H}}{H_0^2}+...,
\end{equation}
based on quantum corrections of QFT in curved spacetime \cite{Sola:2013gha}. The coefficients $\nu$ and $\alpha$ are dimensionless. For $\nu = \alpha = 0$, we recover the cosmological constant. 

The RVM suggests two types of models: type G models where we have running G and hence a running $\rho_{\Lambda}(H)$ (matter remains conserved); type A models where G is constant (with anomalous conservation law). Here we compare the Dynamical Space Time (DST) cosmology with the second type of RVM, that assumes $G = $const. By making this choice we do not loose much generality since, as it is well known by a conformal transformation we can come back to a constant $G$ (in the Einstein frame), and therefore in this proposal we study $G = $const. More general cases will be the subject of our full investigation in the future. This will be related to alternative theories that couple the Einstein term to some scalar fields and give a running Newtonian constant. 

The conservation of the total energy momentum tensor gives the extended Friedmann matter equation:
\begin{equation}\label{eq:HubbleA}
H^2/H_0^2 =\Omega_\Lambda^{(0)}+\frac{\Omega_m^{(0)}}{\xi} a^{-3\xi} +\frac{\Omega_r^{(0)}}{\xi^\prime} a^{-4\xi^\prime}\,.
\end{equation}
The new coupling constant read:
\begin{equation}\label{eq:xixip}
\xi=\frac{1-\nu}{1-\alpha}\equiv 1-\nu_{eff}\,,\ \ \  \xi^\prime=\frac{1-\nu}{1-\frac{4}{3}\alpha}\equiv 1-\nu'_{eff}\,.
\end{equation}
The standard expressions for matter and radiation energy densities are recovered for $\xi,\xi'\to 1$. The lack of an action principle for the RVM may be solved with DST formulation.

 The plan of the work is as follows: In section \ref{sec:DST} we introduce the DST with the complete action and the equations of motion. Section \ref{sec:Diffusive} introduces the diffusive extension to the DST action with the complete solution. Section \ref{sec:Data} confront the model with some data set.  Finally, in section \ref{sec:Con} we summarize our results.
 
\section{Dynamical Space Time Theory}
\label{sec:DST}
\subsection{The Dynamical Time Theory}
The conservation of energy can be derived from the time translation invariance principle. However using a Lagrange multiplier, one can derive the covariant local conservation of an energy momentum tensor $\mathcal{T}^{\mu\nu}$. Let's consider a 4 dimensional case where a conservation of a symmetric energy momentum tensor $\mathcal{T}^{\mu\nu}$ is imposed by introducing the term $\mathcal{S}_{(\chi)}$ in the action \cite{Guendelman:2009ck,Benisty:2018qed,Benisty:2016ybt}:
\begin{eqnarray} \label{action}
\mathcal{S}&=&\frac12 \int d^{4}x\sqrt{-g}\mathcal{R}+ 	\mathcal{S}_{(\chi)} \quad {\rm where} \nonumber \\
	\mathcal{S}_{(\chi)}&=&\int d^{4}x\sqrt{-g} \, \chi_{\mu;\nu} \, \mathcal{T}^{\mu\nu},
\end{eqnarray}
and $ \chi_{\mu;\nu}=\partial_{\nu}\chi_{\mu}-\Gamma_{\mu\nu}^{\lambda}\chi_{\lambda}$. 
 The vector field $\chi_\mu$ called a dynamical space time vector, because the energy density of $T^{\mu\nu}_{(\chi)}$ is a canonically conjugated variable to $\chi_0$, which is what we expected from a dynamical time:
\begin{equation} 
	\pi_{\chi_{0}} = \frac{\partial \mathcal{L}}{\partial \dot{\chi}^0} = T^{0}_{0} (\chi) 
\end{equation}
In the metric formalism, the variation with respect to $ \chi_{\mu} $ gives a covariant conservation law:
\begin{equation}
\nabla_{\mu}\mathcal{T}^{\mu\nu}=0
\end{equation}
The covariant conservation of the $\mathcal{T}^{\mu\nu}$ is satisfied because of the variation with respect to the dynamical spacetime vector field. However, the covariant conservation of the metric energy momentum tensor is $G_{\mu\nu}$ is fulfilled automatically because of the Bianchi identity.

The reason we call this four vector the dynamical space time is because its canonical momentum is an energy density, so as we normally associate time as the conjugate of energy, this seems a natural identification. Furthermore in many solutions the dynamical time coincides with the cosmic time and when this does not happens exotic effects happen. Notice finally that in GR the time is just a coordinate and can be set to anything we want, so it is meaningless to talk about the dynamics of a coordinate, unlike the zero component of a four vector.

A particular case of the stress energy tensor with the form $\mathcal{T}^{\mu\nu}=\mathcal{L}_{1}g^{\mu\nu}$ corresponds to a modified measure theory. By substituting this stress energy tensor into the action itself, the determinant of the metric is cancelled:
\begin{equation}\label{mm}
\sqrt{-g}\chi^{\mu}_{;\mu} \mathcal{L}_{1} 
= \partial_\mu(\sqrt{-g}\chi^{\mu}) \mathcal{L}_{1} = \Phi \mathcal{L}_{1}
\end{equation}
where $\Phi=\partial_\mu(\sqrt{-g}\chi^{\mu})$ is like a ``modified measure''. This situation corresponds to the ``Non-Riemannian Volume-Forms'' \cite{Guendelman:1999qt,Guendelman:1999tb,Comelli:2007id}, where in addition to the regular measure of integration $\sqrt{-g}$, the Lagrangian includes a modified measure of integration, which is also a scalar density and a total derivative. with the modified measure being generalized by using the dynamical space time vector field $\chi_\mu$.

A variation with respect to the dynamical time vector field will give a constraint on $\mathcal{L}_1$ to be a constant:
\begin{equation}\label{constraint}
{\partial_{\alpha}\mathcal{L}_{1}=0} \quad \Rightarrow \quad \mathcal{L}_{1} = Const
\end{equation}

Some basic symmetries that holds for the dynamical space time theory are two independent shift symmetries:
\begin{equation}\label{eq:shift}
\chi_\mu \rightarrow \chi_\mu + k_\mu,\quad \mathcal{T}^{\mu\nu} \rightarrow \mathcal{T}^{\mu\nu} + \Lambda g^{\mu\nu}
\end{equation}
where $\Lambda$ is some arbitrary constant and $k_\mu$ is a Killing vector of the solution.
This transformation does not change the action (\ref{action}) , which means that the redefinition of the energy momentum tensor (\ref{eq:shift}) does not change the equations of motion. Of course such type of redefinition of the energy momentum tensor is exactly what is done in the process of normal ordering in Quantum Field Theory for instance.

\subsection{Running Vacuum with Dynamical Time}
In this section we consider the following action:
\begin{equation}
\mathcal{L} = \frac{1}{2}\mathcal{R} + \chi_{,\mu;\nu}\mathcal{T}^{\mu\nu} -\frac{1}{2}\phi^{,\mu}\phi_{,\mu} - V(\phi)
\end{equation}
which contains a scalar field with potential $V(\phi)$. The stress energy momentum tensor $\mathcal{T}^{\mu\nu}$ is chosen to be:   
\begin{equation}
\mathcal{T}^{\mu\nu} = -\frac{\lambda_1}{2} \phi^{,\mu}\phi^{,\nu} - \frac{\lambda_2}{2} g^{\mu\nu} (\phi_{,\alpha}\phi^{,\alpha})+g^{\mu\nu}U(\phi)
\end{equation} 
where $\lambda_1$ and $\lambda_2$ are arbitrary constants, $U(\phi)$ is another potential. In such a case the density and pressure resulting from $\mathcal{T}^{\mu\nu}$ are:
\begin{equation}\label{11:Tmunu}
\tilde{\rho} = (\lambda_1+\lambda_2)\frac{\dot{\phi}^2}{2} + U(\phi), \quad
\tilde{p} = -\lambda_2\frac{\dot{\phi}^2}{2} - U(\phi)
\end{equation}
with the original energy momentum tensor is: $\mathcal{T}^{\mu}_{\nu} =  (\tilde{\rho},-\tilde{p} ,-\tilde{p} ,-\tilde{p}) $. For simplicity we take $U(\phi) = \text{const}$. Because of the symmetry \eqref{eq:shift}, the $U(\phi)$ does not contribute to the action.

The action depends on three different variables: the scalar field $\phi$, the dynamical space time vector $\chi_{\mu}$ and the metric $g_{\mu\nu}$. Because we assume homogeneous background, the scalar field is assumed to be depend only on time $\phi = \phi(t)$. The vector field is assumed to be in the form:
\begin{equation}
\chi_\mu = (\chi_0(t),0,0,0).
\end{equation}
The metric we use is the Friedmann Lemaitre Robertson Walker Metric (FLRW), with a Lapse function: 
\begin{equation}
ds^2 = -\mathcal{N}(t) dt^2 +a(t)^2 (dx^2 +dy^2 +dz^2),
\end{equation}
 where $a$ is the scale factor and the $\mathcal{N}(t)$ is the Lapse function, which in the equations of motion is gauged to be $\mathcal{N}(t) = 1$. In Mini-Super-Space, the action (\ref{action}) reads: 
\begin{equation}
\begin{split}
\mathcal{L}_{M.S.S}  = \frac{3 a^2 \ddot{a}}{\mathcal{N}}-\frac{3 a^2 \dot{a} \dot{\mathcal{N}}}{\mathcal{N}^2}-\frac{3 \lambda_2 a^2 \chi_0 \dot{a} \dot{\phi}^2}{2 \mathcal{N}^3}+\frac{3 a \dot{a}^2}{\mathcal{N}}+\frac{\lambda_1 a^3 \chi_0 \dot{\mathcal{N}} \dot{\phi}^2}{2 \mathcal{N}^4}+\frac{\lambda_2 a^3 \chi_0 \dot{\mathcal{N}} \dot{\phi}^2}{2 \mathcal{N}^4}\\-\frac{\lambda_1 a^3 \dot{\chi}_0 \dot{\phi}^2}{2 \mathcal{N}^3}-\frac{\lambda_2 a^3 \dot{\chi}_0 \dot{\phi}^2}{2 \mathcal{N}^3}-a^3 \mathcal{N} V(\phi)+\frac{a^3 \dot{\phi}^2}{2 \mathcal{N}} 
\end{split}
\end{equation}
The variation with respect the Dynamical Time vector field $\chi_0$ yields:
\begin{equation}\label{1frw}
\frac{3}{2} \lambda_1 H \dot\phi +\lambda \ddot{\phi}=0
\end{equation}
which is integrated to give:
\begin{equation}
\dot{\phi}= H_0 \sqrt{2\Omega_m^{0}} a^{-3\lambda_1/2\lambda},
\label{11:phi}
\end{equation}
with an integration constant $H_0 \sqrt{2\Omega_m^{0}}$ and $\lambda = \lambda_1 + \lambda_2$. The second variation with respect to the scalar field $\phi$ gives:
\begin{equation}
\begin{split}
2 \lambda  \left(3 (\lambda -\lambda_1) \chi_0 \dot{H}+\lambda  \ddot{\chi}_0\right)+9 H^2 \left(\lambda ^2-\lambda_1^2\right) \chi_0 \\+H \left(3 \lambda  (3 \lambda -\lambda_1) \dot{\chi}_0-3 (\lambda +\lambda_1)\right) = 0  \end{split}
\label{11:chi}
\end{equation}
The last variation, with respect to the metric, gives the ``gravitational'' stress energy tensor, which, as we anticipated, differs from the energy momentum that appears in the action. The energy density and the pressure of the scalar field which are the source of the Einstein tensor are:
\begin{subequations}
\begin{equation}\label{11:den}
\begin{split}
\rho=\frac{1}{2} \dot\phi^2 \left(H (9 \lambda_1-6 \lambda ) \chi_0-2 \lambda  \dot{\chi}_0+1\right)+\lambda  \chi_0 \dot{\phi} \ddot{\phi}  + V(\phi),    
\end{split}
\end{equation}
\begin{equation}
p = (\lambda -\lambda_1) \chi_0 \dot{\phi} \ddot{\phi}+\frac{1}{2} \dot{\phi}^2 \left(\lambda_1 \dot{\chi}_0-1\right)-V(\phi).
\end{equation}
\end{subequations}
with the Friedmann equations:
\begin{equation}
\rho = 3 H^2, \quad  p = -3H^2 - 2 \dot{H}.
\end{equation}
In order to track the evolution of the solution, we use the asymptotic solution: with a power law and an exponential expansion. We assume $V(\phi) = Const = 3H_0^2 \Omega_\Lambda^{0} $ for simplicity.
By taking the first and the second Friedmann equations together we get:
\begin{equation}
\begin{split}
\dot{H} = \frac{3 (3 \lambda_1+2 \lambda_2) \left(H_0^2 \Omega_\Lambda -H^2\right)}{4 (\lambda_1+\lambda_2)} -\frac{H_0^2 \Omega_m^{0} (\lambda_1+2 \lambda_2) H^{-\frac{3 \lambda_1}{\lambda_1+\lambda_2}}}{4 (\lambda_1+\lambda_2)}.
\end{split}
\end{equation}
This equation is independent of $\chi_0$ and it's derivative. From integration we get the extended Friedmann equation:
\begin{equation}
H(t)^2 = H_0^2 \left( \Omega_m^{0} a^{-3\beta} + \Omega_\Lambda \right)
\label{eq:finDST}
\end{equation}
with the density $\rho = 3 H^2$ and the ``matter components'' have a modified power in the Friedmann equations:
\begin{equation}
\beta = \lambda_1/\lambda.
\end{equation}
where $b$ is the multiple modification for the power matter fields. Notice that for the case $\lambda = \lambda_1$ the solution should be different which has been solved analytically and numerically in \cite{Benisty:2018qed,Anagnostopoulos:2019myt}. The RVM energy density (of the second type) corresponds to the asymptotic solution of the DST cosmology for $\xi = b = \lambda_1/\lambda, \xi' = 1$.

Now we have solved the effective matter density without any assumptions concerning $a$, as we consider more complicated situations, in particular,  when we will consider the situation where diffusion
is present, then, in order to study the evolution of the solution for an asymptotically constant solution, we use a power law and exponential expansion forms for $a$. Then we can check that our answers are correct in the limit where the problem has been resolved without any assumptions for $a$. 

If we assume power law solution for the scale factor for large times $a \sim t^{\alpha}$ with an asymptotically constant potential $V=$const. Using power law scale factor in Eq. (\ref{11:chi}), we get the solution for $\chi_0$ as:
\begin{equation}
\chi_0(t)=\frac{t}{\lambda+3\alpha(\lambda-\lambda_1)}+B_1 t^{1-\frac{3\alpha(\lambda+\lambda_1)}{2\lambda}}+B_2t^{-\frac{3\alpha(\lambda-\lambda_1)}{\lambda}}
\end{equation}
where $B_1$ and $B_2$ are integration constants. For large time, considering $2\lambda<3\alpha(\lambda_1+\lambda)$, the second and third terms become sub dominating, hence can be neglected.
Therefore, the solution for $\chi_0$ simplifies to
\begin{equation}\label{11:chisol}
\chi_0(t)=\frac{t}{\lambda+3\alpha(\lambda-\lambda_1)}
\end{equation}
Substituting the solutions for the derivative of $\phi$ (\ref{11:phi}) and the solution of $\chi_0$ from Eq. (\ref{11:chisol}) into the density equation (\ref{11:den}) giving:
\begin{equation}
\rho = C_1^2  \frac{(\lambda+3\alpha(\lambda_1+3\lambda))}{2(3\alpha(\lambda_1-\lambda)+\lambda)}  a^{-3\lambda_1/\lambda}+V.
\end{equation}
We can also obtain the same asymptotic behavior if we consider exponential scale factor given by $a \sim e^{H_0 t}$. Similarly, solution for $\chi_0$ is given by
\begin{equation}
\chi_0(t)=\frac{1}{3H_0(\lambda-\lambda_1)}+C_1e^{-\frac{3tH_0(\lambda+\lambda_1)}{2\lambda}}+C_2e^{-\frac{3tH_0(\lambda-\lambda_1)}{\lambda}}
\end{equation}
Once again for large times, one can neglect the last two terms, hence
\begin{equation}
\chi_0(t)=\frac{1}{3H_0(\lambda-\lambda_1)}
\end{equation}
Substituting the above solutions into the density equation (\ref{11:den}), we get the expression:
\begin{equation}
\rho =  C_1^2\frac{\lambda_1+3\lambda}{2(\lambda_1-\lambda)} a^{-3\lambda_1/\lambda}+V
\end{equation}
which is similar to the power law expansion, but with different coupling constants. 

Notice that for the case $\lambda = \lambda_1$ the solution should be different, but solved analytically and numerically in Ref. \cite{Benisty:2018qed,Anagnostopoulos:2019myt}. The RVM energy density (from the second type) corresponds to the asymptotic solution of the DST cosmology for:
\begin{equation}
\xi = \lambda_1/\lambda, \quad \xi' = 1
\end{equation}
In this paper we will ignore the radiation, although we have studied radiation from gauge fields in cosmology in the context of the DST theory, considering a non trivial coupling of the dynamical space time vector to a radiation energy momentum tensor \cite{Benisty:2016ybt}. The analysis is a bit more complicated. In order to modify the matter part and leave the radiation as an external field, or to say that it does not couple to our dynamical space time vector field. Here in order to obtain both the matter and vacuum energy  densities we have used the DST action principle, or as we will do in the next chapter, we will use the extension of the DST that gives a Diffusive action.

\section{Diffusive Extension}
\label{sec:Diffusive}
\subsection{The Diffusion Theories}
In order to break the conservation of $\mathcal{T}^{\mu\nu}$ as in the diffusion equation, the vector field $\chi_\mu$ a mass like term could be added in the action:
\begin{equation}\label{nhd1}
\begin{split}
S_{(\chi,A)}=\int d^{4}x\sqrt{-g}\chi_{\mu;\nu}\mathcal{T}^{\mu\nu} + \frac{\sigma}{2}\int d^4x \sqrt{-g}(\chi_{\mu}+\partial_{\mu}A)^2 \end{split}
\end{equation} 
where $A$ is a  scalar field different from $\phi$. From a variation with respect to the dynamical space time vector field $\chi_{\mu}$, we obtain:
\begin{equation} \label{nhd2}
\nabla_{\nu}\mathcal{T}^{\mu\nu}=\sigma(\chi^{\mu}+\partial^{\mu}A)= f^\mu,
\end{equation} 
where the current source reads: $f^\mu=\sigma (\chi^{\mu}+\partial^{\mu}A)$. From the variation with respect to the new scalar $A$, a covariant conservation of the current  indeed emerges: 
\begin{equation}\label{nhd3}
\nabla_{\mu}f^\mu=\sigma\nabla_{\mu}(\chi^{\mu}+\partial^{\mu}A)=0
\end{equation} 
A particular case of diffusive energy theories is obtained when $\sigma \to \infty$. In this case, the contribution of the current $f_\mu$ in the equations of motion goes to zero and  yields a constraint for the vector field being a gradient of the scalar:
\begin{equation}
f_\mu=\sigma(\chi_{\mu}+\partial_{\mu}A) =0 \quad \Rightarrow \quad \chi_{\mu}=-\partial_{\mu}A
\end{equation}
The theory (\ref{nhd1}) is reduced to a theory with higher derivatives:
\begin{equation}\label{action2}
\mathcal{S}= - \int d^{4}x\sqrt{-g} \,  A_{,\mu;\nu} \, \,  \mathcal{T}^{\mu\nu}
\end{equation}
The variation with respect to the scalar $A$ gives
\begin{equation}
\nabla_\mu \nabla_\nu \mathcal{T}^{\mu\nu}=0    
\end{equation}
which  corresponds to the variations (\ref{nhd2}) - (\ref{nhd3}). In the following part of this paper we use the reduced theory with higher derivative in the action. 
The covariant conservation of the $\mathcal{T}^{\mu\nu}$ can break because of the variation with respect to the scalar $A$. In such a case, the doubled divergence of the $\mathcal{T}^{\mu\nu}$ is zero. However, the covariant conservation of the metric energy momentum tensor $G_{\mu\nu}$ is fulfilled automatically because of the Bianchi identity.

\subsection{The diffusive extension}
We consider the following action: \cite{Benisty:2017lmt,Bahamonde:2018uao,Benisty:2018oyy}:
\begin{equation}\label{action1}
\mathcal{L} = \frac{1}{2}\mathcal{R} + A_{,\mu;\nu}\mathcal{T}^{\mu\nu} -\frac{1}{2}\phi^{,\mu}\phi_{,\mu} - V(\phi)
\end{equation}
which contains a scalar field with potential $V(\phi)$. There are three independent sets of equations of motions: $A$, $\phi$ and the metric $g_{\mu\nu}$.

In the Mini-Super-Space the action (\ref{action1}) reads:
\begin{equation}
\begin{split}
\mathcal{L}_{M.S.S} = \frac{3 a^2 \ddot{a}}{\mathcal{N}}-\frac{3 a^2 \dot{a} \dot{\mathcal{N}}}{\mathcal{N}^2}-\frac{3 \lambda_2 a^2 \dot{a} \dot{A} \dot{\phi}^2}{2 \mathcal{N}^3}+\frac{3 a \dot{a}^2}{\mathcal{N}}+\frac{\lambda_1 a^3 \dot{\mathcal{N}} \dot{A} \dot{\phi}^2}{2 \mathcal{N}^4}+\frac{\lambda_2 a^3 \dot{\mathcal{N}} \dot{A} \dot{\phi}^2}{2 \mathcal{N}^4}\\-\frac{\lambda_1 a^3 \ddot{A} \dot{\phi}^2}{2 \mathcal{N}^3}-\frac{\lambda_2 a^3 \ddot{A} \dot{\phi}^2}{2 \mathcal{N}^3}-a^3 \mathcal{N} V(\phi)+\frac{a^3 \dot{\phi}^2}{2 \mathcal{N}}   
\end{split}
\end{equation}
Again here we assume $U(\phi) = Const$. According to this ansatz the scalar fields are solely functions of time. The variation with respect to the scalar $A$ gives:
\begin{equation}\label{chisol}
(\lambda_1+\lambda_2)\dot{\phi}\ddot{\phi} + 3 H \lambda_1\dot{\phi}^2 = \frac{\sigma_1}{a^3}
\end{equation}
where $\sigma_1$ is an integration constant. Then the solution for Eq. (\ref{chisol}) is:
\begin{equation}\label{chisol2}
\dot{\phi}^2  = \dot{\phi}_{(0)}^2 a^{-\frac{3\lambda_1}{\lambda_1+\lambda_2}}+ \frac{\sigma_1}{\lambda_1+\lambda_2} a^{-\frac{3\lambda_1}{\lambda_1+\lambda_2}} \int_{0}^{t}ds a^{-\frac{3\lambda_2}{\lambda_1+\lambda_2}} 
\end{equation}
In addition for the same theoretical reason we assume that $V(\phi) = {\rm Const}$. Then variation with respect to the scalar field $\phi$ yields: 
\begin{equation}\label{chi3}
(\lambda_1-\lambda_2) \ddot{A} + (1 - 3H \lambda_2 \dot{A})  = \frac{\sigma_2}{\dot{\phi} a^3}
\end{equation}
where $\sigma_2$ is another integration constant. Now from the stress energy momentum tensor, the total energy density term is:
\begin{equation}\label{density}
\begin{split}
\rho = \frac{3}{2} H (\lambda_1-2 \lambda_2) \dot{A} \dot{\phi}^2+\frac{1}{2} \dot{\phi}^2 \left(1-2 (\lambda_1+\lambda_2) \ddot{A}\right)+\dot{A} \dot{\phi} \left((\lambda_1+\lambda_2) \ddot{\phi}\right)+V,
\end{split}
\end{equation}
and the total pressure is:
\begin{equation}
p = \frac{1}{2} \dot{\phi}^2 -\frac{1}{2}\lambda_1 \ddot{A} \dot\phi^2+\lambda_2 \dot{A} \dot\phi \ddot{\phi}-V.
\end{equation}
We aren't able to find the exact solutions for the Einstein equation together with the equations for the scalar fields. So we are looking for asymptotic solutions. We assume a power law solution for a large time $a \sim t^{\alpha}$. Then from Eq. (\ref{chisol}) the solution for the scalar field $\phi$ derivative is:
\begin{equation}\label{phipower}
\dot{\phi} = \sqrt{\frac{2\sigma_1}{3 \alpha (\lambda_1-\lambda_2)+\lambda_1+\lambda_2}}\, t^{\frac{1}{2}-\frac{3 \alpha }{2}}
\end{equation}
The solution for the derivative of the scalar field $A$ is:
 \begin{equation}\label{chipower}
 \dot{A} = \frac{2\lambda_2}{-6 \alpha  \lambda_2+\lambda_1-2 \lambda_2}\, t.
 \end{equation}
By inserting the solutions (\ref{phipower}) and (\ref{chipower}) into Einstein equation we obtain:
\begin{equation}
\rho = \frac{\alpha_1}{a^3} + \frac{\alpha_2 t}{a^3}+V
\end{equation}
where the constants are:
\begin{equation}
\alpha_1 = \frac{18 \alpha ^2 \lambda_2 (2 \lambda_2-\lambda_1)}{2 (\lambda_1-2\lambda_2 (3 \alpha  +1))}
\end{equation}
\begin{equation}
\alpha_2 = \frac{(6 \alpha +2) \lambda_1 \lambda_2+2 (3 \alpha +1) (\lambda_2-1)\lambda_2+ \lambda_1}{2 (\lambda_1-2 \lambda_2 (3 \alpha +1))}
\end{equation}
For exponential solution, the asymptotic limit reads different. For this we set $a \sim e^{H_0 t}$ in Eq. (\ref{chisol2}). Then we get: 
\begin{equation}\label{scalarExp}
\dot{\phi}^2 = \dot{\phi}_0^2 a^{-\frac{3\lambda_1}{\lambda_1+\lambda_2}} - \sigma_1 H_0 \frac{\lambda_1+\lambda_2}{3\lambda_2}\frac{1}{a^3}
\end{equation}
if we impose $\frac{3\lambda_1}{\lambda_1+\lambda_2} > 0$. Then from Eq. (\ref{chi3}) we get the density as:
\begin{equation}
\begin{split}
\rho = \frac{H_0 \sigma_0 (\lambda_1+\lambda_2)}{3 \lambda_2}\frac{1}{a^3} + V  + \frac{1}{2} \dot{\phi}_0^2  a^{-\frac{3\lambda_1}{\lambda_1+\lambda_2}}
\end{split}
\label{eq:finDiff}
\end{equation}
The physical quantities corresponds to the parameters of the theory via:
\begin{equation}
\Omega_m^{0} = \frac{H_0 \sigma_0 (\lambda_1+\lambda_2)}{3 \lambda_2}, \quad \Omega_\Lambda^{0} = V, 
\end{equation}
and with a new power:
\begin{equation}
\Omega_\gamma^{0} = \frac{1}{2} \dot{\phi}_0^2, \quad \gamma = 3 \beta.
\end{equation}
With this solution, the corresponding energy density for the RVM is  
\begin{equation}
\xi = 1, \quad \xi' = 3\lambda_1/(4\lambda).
\end{equation}
In this case, we modify the radiation part and obtain the matter field from one action. In both cases we can modify one part of the Friedmann matter equation. However to obtain the full RVM energy density from those theories is not possible, but still implies for the direction how to use Lagrange multipliers as the DST cosmology to obtain the RVM.

\begin{table}[t!]
\tabcolsep 5.5pt
\vspace{1mm}
\centering
\begin{tabular}{cccc} \hline \hline
Parameter & DST  & Diffusive & $\Lambda$CDM
\vspace{0.05cm}\\ \hline \hline
$H_0 (km/Sec/Mpc)$ & $69.63 \pm 1.10$ & $69.33 \pm 1.24$ & $70.40 \pm 1.174$ \\
$\Omega_m$ & $0.2780 \pm 0.0214$ &$0.2785 \pm 0.02$ & $0.263 \pm 0.208$ \\
$\Omega_\Lambda$ & $0.703 \pm 0.024$ &$0.704 \pm 0.0203$ & $ 0.728 \pm 0.0151$\\
$\beta $ & $0.941 \pm 0.0568$ & $0.9120 \pm 0.05498$ & - \\
$r_{d  }\, (Mpc)$ & $147.1 \pm 2.45$ &$146.7 \pm 2.64$  & $ 145.5 \pm 2.66$
\\
$\chi^{2}_{\rm min}/Dof$& $0.960$ &$0.972$ & $ 0.957$\\
$AIC $& $267.0$ & $269.9$ & $264.7$\\
%%-----------------------------
\hline\hline
\end{tabular}
\caption[]{\it{Observational constraints and the
corresponding $\chi^{2}_{\rm min}$ for the DST model, the Diffusive extension and $\Lambda$CDM model. Here we set $\Omega_k = 0$.}}
\label{tab:Res}
\end{table}

%--------------------------------------
\section{Cosmological Probes}
\label{sec:Data}
In order to constraint our model, we employ the following data sets: Cosmic Chronometers (CC) exploit the evolution of differential ages of passive galaxies at different redshifts to directly constrain the Hubble parameter \cite{Jimenez:2001gg}. We use uncorrelated 30 CC measurements of $H(z)$ discussed in \cite{Moresco:2012by,Moresco:2012jh,Moresco:2015cya,Moresco:2016mzx}. As \textbf{Standard Candles (SC)} we use $40$ uncorrelated measurements of the Pantheon Type Ia supernova dataset \cite{Scolnic:2017caz} that were collected in \cite{Anagnostopoulos:2020ctz} and also measurements from Quasars \cite{Roberts:2017nkm} and Gamma Ray Bursts \cite{Demianski:2016zxi}. The parameters of the models are adjusted to fit the theoretical $\mu _{i}^{th}$ value of the distance moduli,
\begin{equation}
 \mu=m-M=5\log _{10}((1+z) \cdot D_{M})+\mu _{0},   
\end{equation}
to the observed $\mu _{i}^{obs}$ value. $m$ and $M$ are the apparent and absolute magnitudes and $\mu_{0}=5\log \left( H_{0}^{-1}/Mpc\right) +25$ is the nuisance parameter that has been marginalized. The luminosity distance is defined by, 
\begin{equation} 
D_M=\frac{c}{H_0} \int_0^z\frac{dz'}{E(z')}, 
\end{equation}
where $\Omega_k = 0$ and $D_L = (1+z)\cdot D_M$. In addition, we use the uncorrelated data points from different Baryon Acoustic Oscillations (BAO) collected in \cite{Benisty:2020otr} from \cite{Percival:2009xn,Beutler:2011hx,Busca:2012bu,Anderson:2012sa,Seo:2012xy,Ross:2014qpa,Tojeiro:2014eea,Bautista:2017wwp,deCarvalho:2017xye,Ata:2017dya,Abbott:2017wcz,Molavi:2019mlh}. Studies of the BAO feature in the transverse direction provide a measurement of $D_H(z)/r_d = c/H(z)r_d$, with the comoving angular diameter distance \cite{Hogg:2020ktc,Martinelli:2020hud}. In our database we also use the angular diameter distance $D_A=D_M/(1+z)$ and the $D_V(z)/r_d$, which is a combination of the BAO peak coordinates above:
\begin{equation}
    D_V(z) \equiv [ z D_H(z) D_M^2(z) ]^{1/3}.
\end{equation}
$r_d$ is the sound horizon at the drag epoch and it is discussed in the corresponding section. Finally for very precise  "line-of-sight" (or "radial") observations, BAO can also measure directly the Hubble parameter \cite{Benitez:2008fs}.
 
All of the data set we use in this paper are uncorrelated (including the BAO collection). We use a nested sampler as it is implemented within the open-source packaged $Polychord$ \cite{Handley:2015fda} with the $GetDist$ package \cite{Lewis:2019xzd} to present the results. The prior we choose is with a uniform distribution, where $\Omega_{r} \in [0;1]$, $\Omega_{m}\in[0.;1.]$, $\Omega_{\Lambda}\in[0.;1.]$, $H_0\in [50;100] Km/sec/Mpc$, $r_d\in [130;160] Mpc$. For the additional parameter we take $\beta\in [0.,2.]$.We also compare the Akaike information criteria (AIC) of the two models applied to the data set \cite{AIC1,Liddle:2007fy,Anagnostopoulos:2019miu}.

\section{Results}
\label{sec:res}
\begin{figure*}[t!]
 	\centering
\includegraphics[width=1\textwidth]{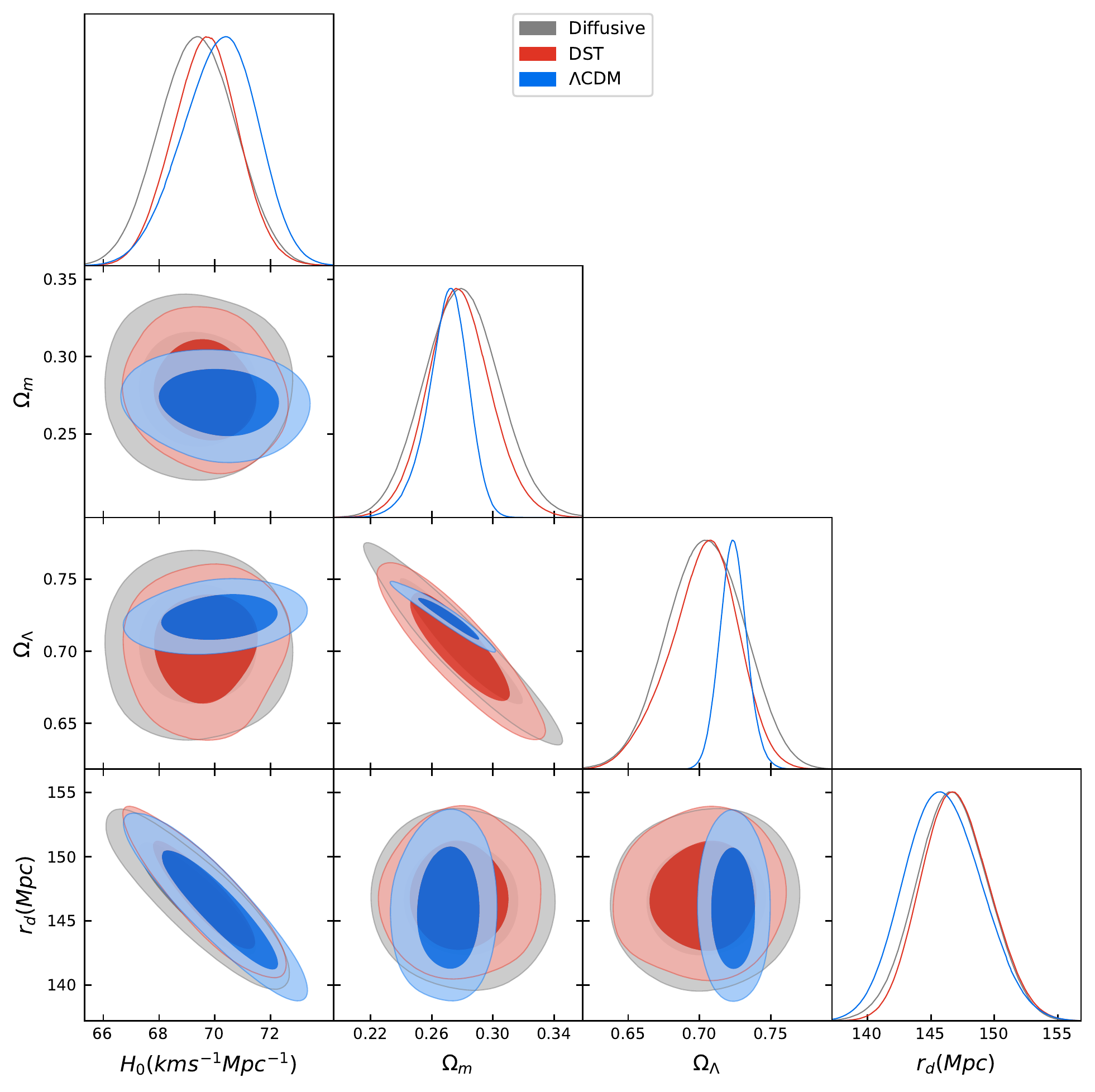}
\caption{\it{The posterior distribution with $1 \sigma$ and $2\sigma$ for the DST, Diffusive and the $\Lambda$CDM model. The data set includes CC + Type Ia Supernova + Quasars + GRB + BAO. The Hubble function for the DST is from Eq. \ref{eq:finDST} and the Hubble function for the Diffusive model is from Eq. \ref{eq:finDiff}.}}
 	\label{fig:res}
\end{figure*}
\begin{figure*}[t!]
 	\centering
\includegraphics[width=1\textwidth]{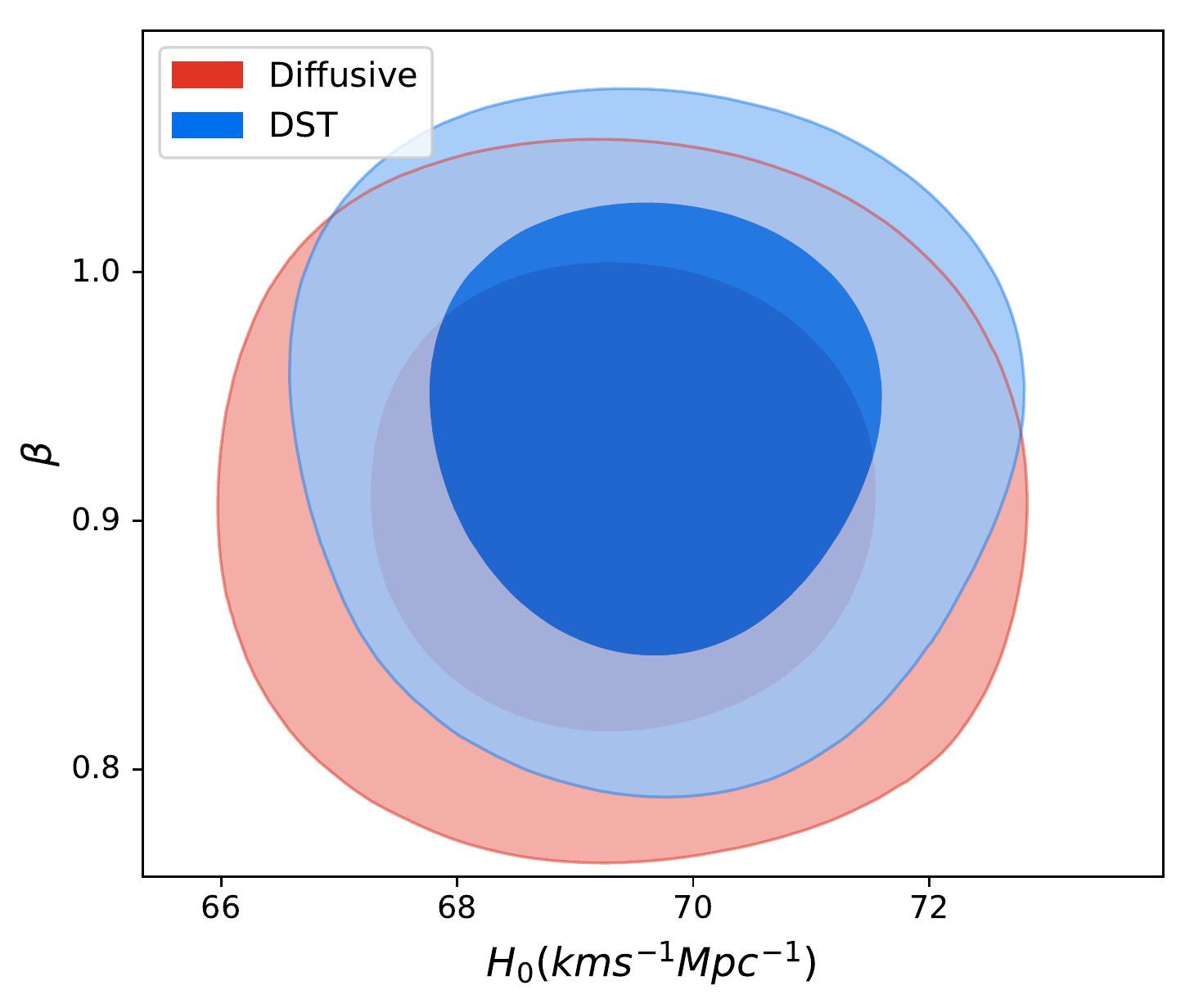}
\caption{\it{The posterior distribution for different measurements with $1 \sigma$ and $2\sigma$ for the Hubble parameter.}}
 	\label{fig:resh0}
\end{figure*}

 
Table \ref{tab:Res} summarizes the results for the RMV and the Diffusive model with a comparison to $\Lambda$CDM model. The posterior distribution is presented in figure \ref{fig:res}, and the corresponding relation between the Hubble parameter and the additional parameter $\beta$ is presented in fig \ref{fig:resh0}. The $\Omega_m$ matter part in the DST model $0.2780 \pm 0.0214$ which is close for the Diffusive case. For the $\Lambda$CDM case the matter part is $0.263 \pm 0.208$. The dark energy $\Omega_\Lambda$ part for the DST case $0.703 \pm 0.024$, for the Diffusive case $0.704 \pm 0.0203$ and for the $\Lambda$CDM case $0.728 \pm 0.009$. The Hubble parameter for the DST case is lower then other cases:  $69.63 \pm 1.10$ for DST, $69.33 \pm 1.24$ for Diffusive case of  $70.40 \pm 1.174 \, (km/Sec/Mpc)$ for $\Lambda$CDM model.

The BAO scale is set by the redshift at the drag epoch $z_d \approx 1020$ when photons and baryons decouple \cite{Aubourg:2014yra}. For a flat $\Lambda$CDM, the Planck measurements yield $147.09 \pm 0.26 Mpc$ and the WMAP fit gives $152.99 \pm 0.97 Mpc$ \cite{Aghanim:2018eyx}. Final measurements from the completed SDSS lineage of experiments in large-scale structure provide $r_d = 149.3 \pm 2.8 Mpc$ \cite{Alam:2020sor}. 

The $\Lambda$CDM model for the combined data set we use gives $ 145.5 \pm 2.66$. However, the DST model gives $147.1 \pm 2.45$, while for the Diffusive case we obtain $146.7 \pm 2.64$. The $r_d$ is in a moderate and reasonable range.
From the AIC we see that $\Lambda$CDM is still the better fit to the late universe, since the AIC for $\Lambda$CDM model $264.7$ is lower then the DST case  $267.0$ or for the Diffusive model case $269.9$.

\section{Discussion}
\label{sec:Con}
We know though $\Lambda$CDM could be the simplest phenomenological explanation for the observed acceleration of the Universe, there still exist a disagreement between the predicted and observed value of $\Lambda$. In particular, we are still facing the crucial question whether $\Lambda$ is truly a fundamental constant or a mildly evolving dynamical variable. It turns out that the $\Lambda=$const, despite being the simplest, may well not be the most favored one when compared with specific dynamical models of the vacuum energy. It also is unable to solve the tension related to the Hubble constant. Recently it has been shown the RVM are good modified models candidates to solve the Hubble tension. However, the model does not consider a Lagrangian formulation directly. 

We obtain a candidate for the RVM formulated by an action principle that approaches the RVM of the second type asymptotically,  without the requirement of dark components. We study this in dynamical space time vector model and also its diffusive extension. The scalar field model takes care of the behavior of the dark components. The kinetic term mimics the behavior of the dark matter and the potential terms acts like dark energy. Because of conformal invariance, the radiation cannot deviate from $1/a^4$ in this scenario. Analysing asymptotically, we have found that the DST and its diffusive counterpart have a different correspondence to the RVM. 

We use CC + Type Ia Supernova + Quasars + GRB + BAO sets in order to constraint the model. It seems from the AIC test that the $\Lambda$CDM model is a better fit then DST and also the Diffusive case. However we can recover $\Lambda$CDM model by setting $\lambda_2 = 0$, and the $\Lambda$CDM in this case is formulated via an action principle.

\section*{Acknowledgments}
SB would like to acknowledge funding from Israel Science Foundation and John and Robert Arnow Chair of Theoretical Astrophysics. DB thanks for Ben Gurion University of the Negev and Frankfurt Institute for Advanced Studies for a great support. EG thanks for Ben Gurion University of the Negev for a great support. This article is supported by COST Action CA15117 "Cosmology and Astrophysics Network for Theoretical Advances and Training Action" (CANTATA) of the COST (European Cooperation in Science and Technology). This project is partially supported by COST Actions CA16104 and CA18108. We want to thank Professor Joan Sola for conversations on the Running Vacuum Model.

\bibliographystyle{unsrt}
\bibliography{ref}

\end{document}